\documentclass[conference]{IEEEtran}
\pdfoutput=1
\usepackage{trackchanges}
\addeditor{Jenny}
\addeditor{Zahra}

\ifCLASSINFOpdf
\else

\fi

\newcommand{\PRLsep}{\noindent\makebox[\linewidth]{\resizebox{0.3333\linewidth}{1pt}{$\bullet$}}\bigskip}
\usepackage[font= scriptsize, skip=0pt]{caption}

\usepackage{booktabs}
\usepackage{enumitem}
\usepackage{varwidth}
\usepackage{capt-of}
\usepackage{listings}
\usepackage{amsmath}
\usepackage{framed}
\usepackage{tabu}
\usepackage{float}
\usepackage[caption = false]{subfig}
\usepackage{enumitem}
\usepackage{color}
\usepackage{graphicx}
\usepackage{cleveref}
\usepackage{fancybox}
\usepackage[export]{adjustbox}
\usepackage[table]{xcolor}
\definecolor{Gray}{gray}{0.9}
\definecolor{White}{RGB}{255,255,255}
\usepackage{bbold}
\usepackage{bbm}
\usepackage{mathbbol}
\usepackage{multirow}
\usepackage{fix-cm}
\usepackage{siunitx,booktabs}
\usepackage{xcolor}

\usepackage{setspace}
\usepackage{graphicx}
\usepackage{colortbl}
\usepackage{array}
\usepackage{pifont}
\usepackage {rotating}
\usepackage{mwe}
\usepackage{pbox}
\usepackage{enumitem}
\let\oldding\ding% Store old \ding in \oldding
\renewcommand{\ding}[2][1]{\scalebox{#1}{\oldding{#2}}}
\usepackage{cite}

\newcommand{\Csharp}{%
  {\settoheight{\dimen0}{C}C\kern-.05em \resizebox{!}{\dimen0}{\raisebox{\depth}{\#}}}}

 \usepackage{enumitem}
\usepackage{xspace}
\usepackage{xparse}
\DeclareDocumentCommand\newstep{o}{%
\item\IfNoValueTF{#1}{}{#1 \textendash\xspace}}

\newlist{steps}{enumerate}{1}
\setlist[steps]{label=\textit{Step \arabic*:},leftmargin=*}

\definecolor{orange}{RGB}{0,32,96}
\definecolor{o}{RGB}{245,245,245}
\definecolor{g}{RGB}{50,50,50}

\PassOptionsToPackage{gray}{xcolor}
\usepackage{tikz}
\newcommand\mybox[2][]{\tikz[overlay]\node[fill=gray,inner sep=1pt, anchor=text, rectangle, rounded corners=0.5mm,#1] {#2};\phantom{#2}}

\usepackage{tcolorbox}

\PassOptionsToPackage{gray}{xcolor}

\begin{document}
\title{\huge What Works Better? A Study of Classifying Requirements}
%\title{\huge What Works Better to Classify Requirements?  }

%\author{\IEEEauthorblockN{Zahra Shakeri Hossein Abad, Guenther Ruhe}
%\IEEEauthorblockA{Department of Computer Science\\ University of Calgary, Calgary, Canada\\
%Email: \{zshakeri, ruhe\}@ucalgary.ca}
%\and
%\IEEEauthorblockN{Oliver Karras, Kurt Schneider}
%\IEEEauthorblockA{Software Engineering Group\\
%Leibniz Universit�t Hannover, Hannover, Germany\\
%Email: \{oliver.karras, kurt.schneider\}@inf.uni-hannover.de}
%
%\IEEEauthorblockN{Parisa Ghazi, Kurt Schneider}
%\IEEEauthorblockA{Software Engineering Group\\
%Leibniz Universit�t Hannover, Hannover, Germany\\
%Email: \{oliver.karras, kurt.schneider\}@inf.uni-hannover.de}}

\author{
    \IEEEauthorblockN{Zahra Shakeri Hossein Abad\IEEEauthorrefmark{1}, Oliver Karras\IEEEauthorrefmark{2}, Parisa Ghazi\IEEEauthorrefmark{3}, Martin Glinz\IEEEauthorrefmark{3}, Guenther Ruhe\IEEEauthorrefmark{1}, Kurt Schneider\IEEEauthorrefmark{2}}
    \IEEEauthorblockA{\IEEEauthorrefmark{1}SEDS Lab, Department of Computer Science, University of Calgary, Calgary, Canada
    \\\{zshakeri, ruhe\}@ucalgary.ca}
    \IEEEauthorblockA{\IEEEauthorrefmark{2}Software Engineering Group, Leibniz Universit\"at Hannover, Hannover, Germany
    \\\{oliver.karras, kurt.schneider\}@inf.uni-hannover.de}
    \IEEEauthorblockA{\IEEEauthorrefmark{3}Department of Informatics, University of Zurich, Zurich, Switzerland
    \\\{ghazi, glinz\}@ifi.uzh.ch}
}

\maketitle

\begin{abstract}
Classifying requirements into functional requirements (FR) and non-functional ones (NFR) is an important task in requirements engineering. However, automated classification of requirements written in natural language is not straightforward, due to the variability of natural language and the absence of a controlled vocabulary.
This paper investigates how automated classification of requirements into FR and NFR can be improved and how well several machine learning approaches work in this context. We contribute an approach for preprocessing requirements that standardizes and normalizes requirements before applying classification algorithms. Further, we report on how well several existing machine learning methods perform for automated classification of NFRs into sub-categories such as usability, availability, or performance. Our study is performed on 625 requirements provided by the OpenScience tera-PROMISE repository.
We found that our preprocessing improved the performance of an existing classification method. We further found significant differences in the performance of approaches such as Latent Dirichlet Allocation,  Biterm Topic Modeling, or Na\"{i}ve Bayes for the sub-classification of NFRs.

%Classifying functional and Non-Functional Requirements (NFRs) at the early stages of the software process yields less refactoring and change in the later stages. However, this task is not always straightforward. The variability of the natural language and the absence of a controlled vocabulary for representing software requirements makes interpreting and automatic classification of the requirements difficult. This paper describes and compares several machine learning approaches for automating the process of detecting functional and non-functional requirements and classifying NFRs to attributes such as usability, availability, and performance. Our proposed approach for preprocessing Software Requirements Specification (SRS) documents is first introduced, which leverages the rich sentence features to reduce the inconsistency inherent in SRS documents. The C4.5 decision tree learning algorithm  is then applied to detect functional and non-functional requirements.   To classify NFRs several classification methods such as Hierarchical, K-means, and Hybrid clustering techniques, along with Binarized Naive Bayes and two variations of Topic Modeling approaches: Latent Dirichlet Allocation (LDA) and Biterm Topic Modeling (BTM) are used and evaluated through reporting an experiment on 625 requirements provided by the OpenScience tera-PROMISE repository. The evaluation results show that the proposed preprocessing technique significantly improved the performance of the classification methods, which can be used as a pre-analysis step for quickly classifying large SRS documents. 

\end{abstract}

\begin{IEEEkeywords}
	Functional and Non-Functional Requirements, Classification, Topic Modeling, Clustering, Na\"{i}ve Bayes
\end{IEEEkeywords}

\IEEEpeerreviewmaketitle

\section{Introduction}
 
 In requirements engineering, classifying the requirements of a system by their kind into \emph{functional requirements}, \emph{quality requirements} and \emph{constraints} (the latter two usually called \emph{non-functional requirements}) \cite{re_glossary} is a widely accepted standard practice today.
 
While the different kinds of requirements are known and well-described today~\cite{Martin}, automated classification of requirements written in natural language into functional requirements (FRs) and the various sub-categories of non-functional requirements (NFRs) is still a challenge \cite{Ernst2010}. This is particularly due to the fact that stakeholders, as well as requirements engineers, use different terminologies and sentence structures to describe the same kind of requirements \cite{RO, IWSPM}. The high level of inconsistency in documenting requirements makes automated classification more complicated and therefore error-prone.

In this paper, we investigate how automated classification algorithms for requirements can be improved and how well some of the frequently used machine learning approaches work in this context. We make two contributions. (1)~We investigate if and to which extent an existing decision tree learning algorithm~\cite{hussain2008using} for classifying requirements into FRs and NFRs can be improved by preprocessing the requirements with a set of rules for (automated) standardizing and normalizing the requirements found in a requirements specification. (2)~We study how well several existing machine learning methods perform for automated classification of NFRs into sub-categories such as availability, security, or usability.

With this work, we address the \emph{RE Data Challenge} posed by the 25th IEEE International Requirements Engineering Conference (RE'17).

 \section{Related Work}
 \label{sec:RW}
 Software Requirements Specifications (SRSs) are written in natural language, with mixed statements of functional and non-functional requirements. There is a growing body of research studies that compare the effect of using manual and automatic approaches for classification of requirements~\cite{nanniu,Ernst2010}. An efficient classification enables focused communication and prioritization of requirements~\cite{janerej2}. Categorization of requirements allows filtering relevant requirements for a given important aspect. Our work is also closely related to the research on automatic classification of textual requirements.

Knauss and Ott \cite{knaussREFSQ} introduced a model of a socio-technical system for requirements classification. They evaluated their model in an industrial setting with a team of ten practitioners by comparing a manual, a semi-automatic, and a fully-automatic approach of requirements classification. 
%and evaluated the performance of their approaches by measuring effort and accuracy of automatic classification recommendations, combining performance of user and tool, and capturing the opinion of the expert-participants in a questionnaire. 
They reported that a semi-automatic approach offers the best ratio of quality and effort as well as the best learning performance. Therefore, it is the most promising approach of the three.

%Their results show that the semi automatic approach offers an advantage compared to the manual approach and offers the best ratio of quality and effort and the best learning performance.
% , such as security, performance, and usability requirements
%Cleland-Huang et al. have investigated mining large requirements documents for non-functional requirements (NFR) (quality requirements)~\cite{janere}, \cite{janerej}. Their results indicates that NFR-Classifier adequately distinguish several types of NFRs. However, further work is needed to improve the results for some other NFR types such as ‘look-and-feel'. Although their study is similar to ours as they trained a classifier to recognize a set of weighted indicator terms, indicative of each type of requirement, but we used different classification algorithms such as the hybrid approach and assessed their  precisions and finally compared their performance with each other. 
Cleland-Huang et al. \cite{janere} investigated mining large requirements documents for non-functional requirements. Their results indicate that NFR-Classifier adequately distinguishes several types of NFRs. However, further work is needed to improve the results for some other NFR types such as 'look-and-feel'. Although their study is similar to ours as they trained a classifier to recognize a set of weighted indicator terms, indicative of each type of requirement, we used different classification algorithms and additionally assessed their precisions and recalls to compare their performance with each other.
%Their approach is similar to ours, as they trained a classifier to recognize a set of weighted indicator terms, indicative of each type of requirement. 
%They obtained a recall of 80\% and a precision of 57\% for the security NFR, but could not find a reliable source of keywords for other NFRs. Instead, they developed a supervised classifier by using human experts to identify an NFR training set. Our research differs because we use different algorithms to classify functional and non-functional requirements, assessed their  precision and finally compared the perfomance with each other. 

Rahimi et al. \cite{monaRE} present a set of machine learning and data mining methods for automatically extracting quality concerns from requirements, feature requests, and online forums. Then, they generate a basic goal model from the requirements specification. Each concern is modeled as a softgoal. For attaching topics to softgoals they used an LDA approach to estimate the similarity between each requirement and the discovered topics. In addition, they used LDA to identify the best sub-goal placement for each of the unattached requirements. However, in this research, we used LDA as one of our approaches for classifying the non-functional requirements. 

%Knauss et al. \cite{KnaussRE} used a Naive Bayes classifier for automatic classification of clarification events in requirement discussion. In our study, Naive Bayes method had the highest performance in detecting 92\% of NFRs correctly among the  applied  machine learning algorithm we applied.

Na\"{i}ve  Bayes  classifier is used in several studies\cite{KnaussRE, koj} for  automatic classification of requirements. Therefore, we included Na\"{i}ve Bayes in our study to be comparable with other classifiers.
\section{The Challenge and Research Questions}
\label{sec:RQ}

\subsection{Context and Data Set}
The challenge put forward by the Data Track of RE'17 consists of taking a given data set and performing an automated RE task on the data such as tracing, identifying/classifying requirements or extracting knowledge. For this paper, we chose the task of automated classification of requirements.

The data set given for this task comes from the OpenScience tera-PROMISE repository\footnote{https://terapromise.csc.ncsu.edu/!/\#repo/view/head/requirements/nfr}. It consists of 625 labeled natural language requirements (255 FRs and 370 NFRs). The labels classify the requirements first into FR and NFR. Within the latter category, eleven sub-categories are defined: (a)~ten \emph{quality requirement categories}: Availability (A), Look \& Feel (LF), Maintainability (MN), Operability (O), Performance (PE), Scalability (SC), Security (SE), Usability (US), Fault Tolerance (FT), and Portability (PO); (b)~one \emph{constraint category}: Legal \& Licensing (L). These labels constitute the ground truth for our investigations.

\subsection{Research Questions}
We frame the goal of our study in two research questions:
 {\it RQ1. How do grammatical, temporal and sentimental characteristics of a sentence affect the accuracy of classifying requirements into functional and non-functional ones?}
 
With this research question, we investigate whether our preprocessing approach, which addresses the aforementioned characteristics, has a positive impact on the classification into FRs and NFRs in terms of precision and recall.

 {\it RQ2. To what extent is the performance of classifying NFRs into sub-categories influenced by the chosen machine learning classification method?}
 % MG: Here is a potential alternative formulation of RQ ":
%{\bf RQ2: }How well do different machine learning classification algorithms perform when classifying non-functional requirements into sub-categories?

With this research question, we study the effects of the chosen machine learning method on the precision and recall achieved when classifying the NFRs in the given data set into the sub-categories defined in the data set.

%\begin{itemize}

% {\it RQ1. }How do grammatical, temporal and sentimental characteristics of a sentence affect the accuracy of classifying software requirements?
 
 %{\bf RQ2: }What machine learning classification method works better to classify NFRs?
 
% {\it RQ2. }To what extent the performance of classifying NFRs is influenced by the chosen machine learning classification method?

%\end{itemize}
\section{Preprocessing of Requirements Specifications}
\label{sec:PP}
In this section, we describe the preprocessing %approach
we applied to reduce the inconsistency of requirements specifications by leveraging rich sentence features and latent co-occurrence relations.
\subsection{Part Of Speech (POS) Tagging} We used the part-of-speech tagger of the Stanford Parser \cite{Klein:2003:AUP:1075096.1075150} to assign parts of speech, such as noun, verb, adjective, etc. to each word in each requirement. 
The POS tags\footnote{Check  https://gist.github.com/nlothian/9240750 for a complete list of tags} are necessary to perform the FR/NFR classification based on the approach of Hussain et al. \cite{hussain2008using}.
\subsection{Entity Tagging}
To improve the generalization of input requirements, we used a ``supervised training data'' method in which all context-based products and users are blinded by assigning names as {\it PRODUCT} and {\it USER}, respectively. To this end, we used the LingPipe NLP toolkit\footnote{http://alias-i.com/lingpipe/} and created the \(SRS\_dictionary\) by defining project specific users/customers and products  (e.g., program administrators, nursing staff members, realtor, or card member marked as USER), such as below:
%To improve the generalization of input requirements, we used a ``supervised training data'' method in which all context based products and users are blinded by assigning names as {\it PRODUCT} and {\it USER}, respectively. To this end, we used LingPipe NLP toolkit \footnote{http://alias-i.com/lingpipe/} and created the \(SRS\_dictionary\) by defining project specific users/customers and products  (e.g., Program Administrators, Nursing Staff Members, Realtor, washing process, and card member marked as USER), such as below:
\vspace{-2mm}
\begin{figure}[H]
%\begin{framed}
\centering
{\includegraphics[scale=0.85]{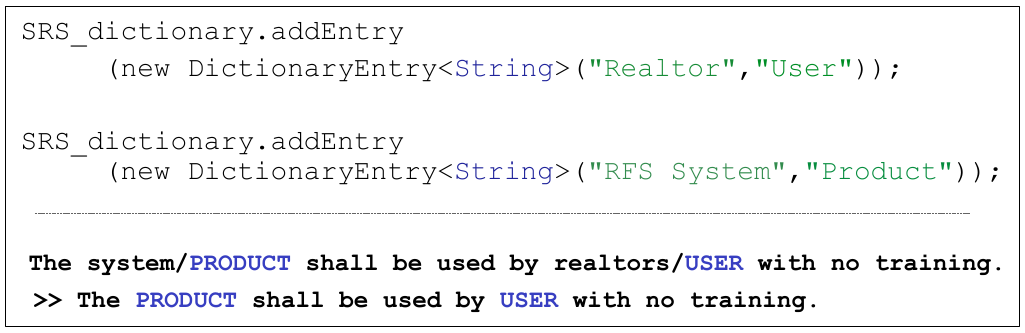}}

\caption*{}

%\end{framed}
\end{figure}

Then, each sentence is tokenized and POS tagged with the developed SRS dictionary. All of the tokens associated with {\it user} and {\it product} were discarded and we only kept these two keywords to represent these two entities. Finally, we used the POS tagger of the Stanford Parser \cite{Klein:2003:AUP:1075096.1075150} and replaced all Noun Phrases (NPs) including ``USER'' and ``PRODUCT'' with \(USER\) and \(PRODUCT\), respectively. For instance, \mybox[fill=gray!30]{registered USER} in \big<Only registered {\bf USER} shall be able to access the {\bf PRODUCT}\big>, is replaced with \(USER\). 
\subsection{Temporal Tagging} {\it Time} is a key factor to characterize non-functional requirements, such as availability, fault tolerance, and performance. 

For this purpose, we used  SUTime, a rule-based temporal tagger for recognizing and normalizing temporal expressions by TIMEX3 standard\footnote{See http://www.timeml.org for details on the TIMEX3 tag}. SUTIME detects the following basic types of temporal objects \cite{SUTIME}:
\begin{enumerate}
\item {\it Time:}  A particular instance on a time scale. SUTIME also handles absolute times, such as {\it Date}. As in: 

\begin{figure}[H]
%\begin{framed}
\centering
{\includegraphics[scale=0.67]{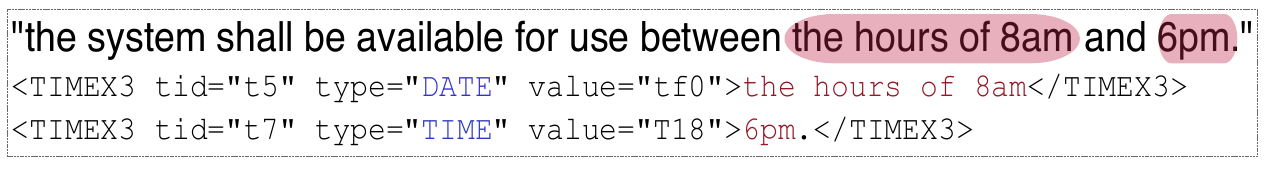}}
%\vspace{-5mm}

\caption*{}

\end{figure}

\item {\it Duration and Intervals:} The amount of intervening time in a time interval. As in:
%\item {\bf Duration and Intervals:} The amount of intervening time between the two end-points of a time interval. As in:

\begin{figure}[H]
%\begin{framed}
\centering
{\includegraphics[scale=0.67]{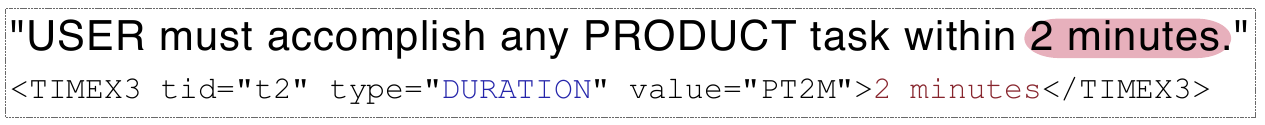}}

\caption*{}

\end{figure}

Intervals can be described as a range of time defined by a start and end time points. SUTime represents this type in the form of other types. 
\item{\it Set:} A set of temporals, representing times that occur with some frequency. As in:

\begin{figure}[H]
%\begin{framed}
\centering
{\includegraphics[scale=0.67]{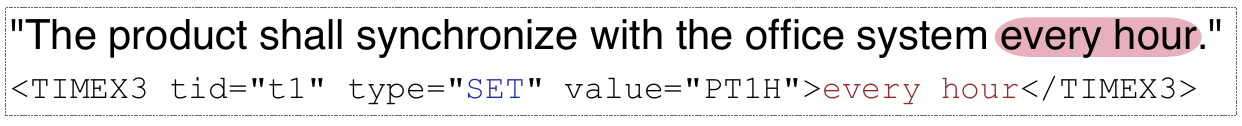}}

\caption*{}
%\end{framed}
\end{figure}
\end{enumerate}

\begin{table*}[!htb]
\centering
\caption{\small Proposed co-occurrence and regular expressions for preprocessing SRSs [(\(CO(w)\)): the set of terms co-occur with word \(w\)] }

\label{tab:reg}
\begin{tabular}{ |c|p{4.1cm}|p{5.1cm}p{5.2cm}| }
  \hline
 {\bf NFR}&{\bf Keywords} & {\bf Part of Speech (POS) and Regular Expressions} & {\bf Replacements}\\\hline 
{\bf Security [SE]} &protect, encrypt, policy, authenticate, prevent, malicious, login, logon, password, authorize, secure, ensure, access&
\( (\mybox[fill=gray!30]{only}/ .../\mybox[fill=gray!30]{ nsubj}/.../\mybox[fill=gray!30]{root}/... )\)
\(  \mid\mid (\mybox[fill=gray!30]{only}/ .../\mybox[fill=gray!30]{ root}/.../\mybox[fill=gray!30]{nmod: agent}/... )\)
\(\wedge \big(\mybox[fill=gray!30]{root}\text{ is a } VBP\big)\)

& \( \Rightarrow \begin{cases}\mybox[fill=gray!30]{nsubj}\mybox[fill=gray!30]{agent} \gets\text{\it authorized user}&\\ \mybox[fill=gray!30]{root} \gets access& \end{cases}\)

{\raisebox{-\totalheight}{\includegraphics[scale=.43]{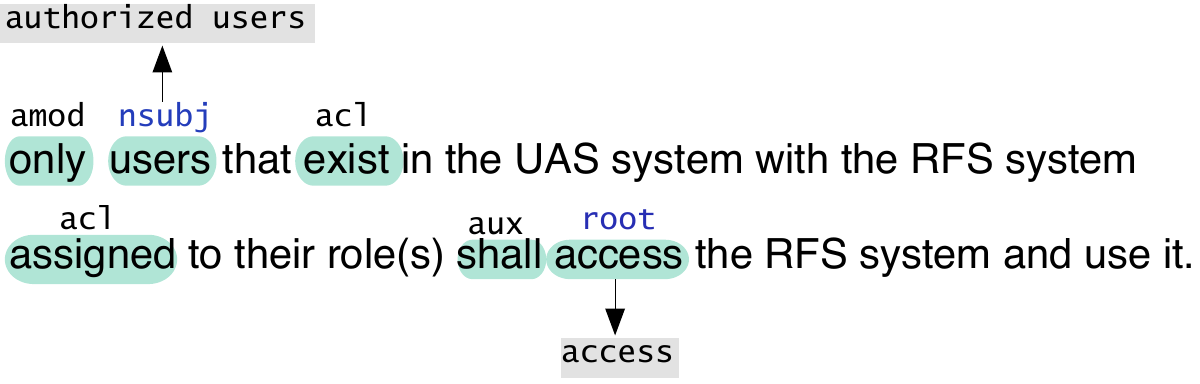}}}\\\cline{3-4}

&&\( \forall \omega \in \{\text{\it username \& password, login, logon}\}\)

 \(\text{\it security, privacy, right, integrity, polict}\}\)
  
  \(\wedge CO(\omega)\cap CO(NFR_{se}) \neq \emptyset\)
  &\(\Rightarrow \omega \gets authorization\)\\\cline{3-4}
  
  &&\( \forall \omega \in \{\text{\it reach, enter, protect, input, interface}\)
  
  \(\wedge \text{product is obj }\wedge CO(\omega)\cap CO(NFR_{se}) \neq \emptyset\)
  &\(\Rightarrow \omega \gets access\)\\\hline

   \hline
\end{tabular}
\vspace{-6mm}
\end{table*}
\vspace{-6mm}

After tagging a sample set of all the classified NFRs, we identified the following patterns and used them to normalize the entire data set.  To do this, first, we replaced all expressions in this format ``24[-//*]7'' and ``\(24\times7\times365\)'' with ``24 hours per day 365 days per year'', and ``everyday'' with ``every day''. Likewise, we replaced all ``sec(s)'' and ``min(s)'' with  ``seconds'' and  ``minutes'', respectively. In the rest of this section, we present the rules we defined and applied to normalize the temporal expressions embedded in requirements' descriptions. \\
\vspace{-3mm}
\begin{tcolorbox}[colback=white, title= Temporal Rules]
\scriptsize
\begin{enumerate}
\item \(\forall \text{   } [\backslash exp]\big<DURATION\big>, exp \gets within\) \\
where \(exp \in\)\{\it no longer than, under, no more than, not be more than, no later, in, for less than, at a maximum\} \\
\vspace{-3mm}

 \vspace{2mm}
\item \(\forall \text{   } [\backslash DURATION \backslash TIME \backslash DATE]^+  \gets alltimes\) \\

\begin{minipage}[t]{1\linewidth}
     \includegraphics[scale=.55]{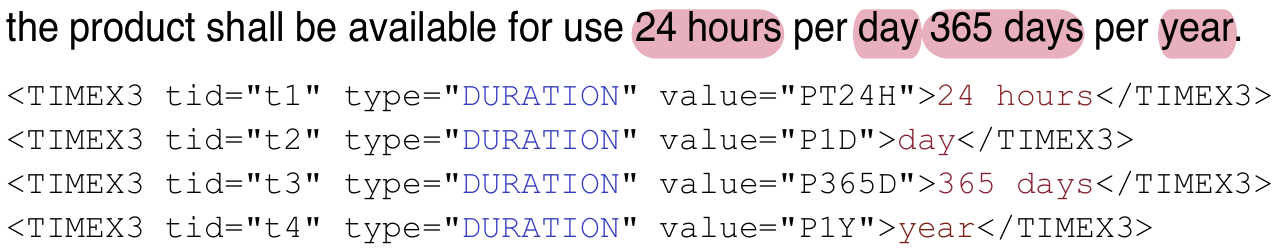}\hfill
        \vspace{-7mm}
        \captionsetup{labelformat=empty}
     { \captionof{figure}{ }
 \label{fig:Terminology}}
    \end{minipage}\hfill\\
    \vspace{2mm}

\item  \(within \text{ } \big<DURATION\big> \gets fast\) \\
\(if \big<DURATION\big> == [\backslash seconds \backslash minutes]\)\\
 \vspace{-2mm}
 
\begin{minipage}[t]{1\linewidth}
\vspace{-2mm}
     \includegraphics[scale=.6]{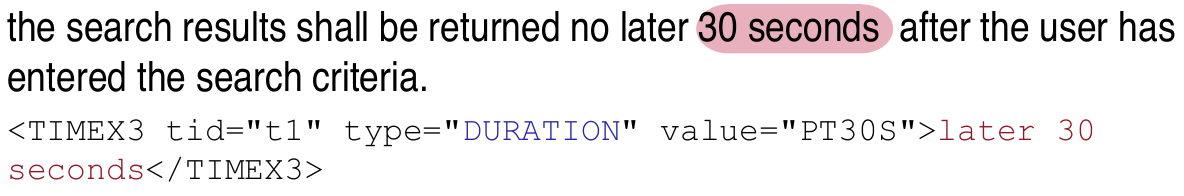}
        \vspace{-3mm}
        \captionsetup{labelformat=empty}
     { \captionof{figure}{ }
 \label{fig:Terminology}}
    \end{minipage}\hfill\\
\vspace{-5mm}

 \vspace{2mm}
\item \(\{timely, quick\} \mid\mid [\backslash \text{\it positive adj } \backslash time] \gets fast\)\\
\item \({[8\text{-}9][0\text{-}9][{\bf \backslash.}?[0\text{-}9]?{\bf \%}?]}[IN\mid DET]^* time \gets alltimes\)

\vspace{-1mm}
 \noindent   

\end{enumerate}
\end{tcolorbox}
\vspace{-5mm}

%\begin{figure}
%%\begin{framed}
%\centering
%\subfloat[]{\includegraphics[scale=0.7]{Figures/SE}}\\[-1mm]
%\subfloat[]{\includegraphics[scale=0.7]{Figures/SE2}}
%\caption{An example for the application of 'Universal dependencies' for processing Security Requirements. }
%%\end{framed}
%\end{figure}

%\begin{algorithm}
%\small
% \KwIn{term \(t_{new} \in\) high frequent words of  \(Corpus_{NFR_i}\), term \(t_{old}\)}
% \KwOut{updated \(Corpus_{NFR_i}\)}
% \Begin{
%parse each requirement using Stanford Parser
% {
% 
%  \ForEach {\(term\) \(t \in T_{old}\)} 
%  {
%  \If{\(POS_{new}== POS_{old}\) and \(t_{new} \in WordNet(t)\)}{
%     \ForEach {\(\big<t + \{CO_{NFR_i}\}\big>^+\)}{
%     \(\)\\
%  \(t \gets t_{new}\)
%   }
%   
%   }
%
%  {\it update the corpus for \(NFR_i\)}
%  }
%  
%   \Return{\(Corpus_{NFR_i}\)}
% }
% }
%{ \caption{\small Replacement \cite{algorithm} for preparing the dataset for analysis}}
% \label{alg:alg}
%\end{algorithm}

\subsection{Co-occurrence and Regular Expressions}
Once the sentence features are utilized to reduce the complexity of the text, we used the co-occurrence and regular expressions to increase the weight of the influential words for each type of NFRs. To explore these rules we manually analyzed 6 to 10 requirements of each NFR and deployed different components of Stanford Parser such as part-of-speech, named entities, sentiment, and relations. Moreover, in this step, we recorded co-occurrence counts of each term within the provided NFR data set as the co-occurrence vector. We used this parameter as a supplement for exploring the SRS regular expressions. For instance, Table \ref{tab:reg} represents the details of the rules we proposed for Security (SE) NFR. Please refer to the footnote\footnote{ http://wcm.ucalgary.ca/zshakeri/projects} for the complete list of these rules containing regular expressions for all of the provided NFRs.

\section{Analysis and Results}
\label{sec:results}
\subsection{RQ1- Influence of grammatical,  temporal  and  sentimental   sentence  characteristics on  FR/NFR classification}
For the classification of functional and non-functional requirements, we used the approach of Hussain et al. \cite{hussain2008using}. We applied this approach to the unprocessed data set of requirements as well as on the processed one resulting from our preprocessing.

{\it Classification Process:}
Firstly, we clean up the respective data set by iteratively removing encoding and formatting errors to ensure the further processing. Subsequently, we apply the part-of-speech tagger of the Stanford Parser \cite{Klein:2003:AUP:1075096.1075150} to assign parts of speech to each word in each requirement.

Based on the tagging of all requirements, we extract the five syntactic features \textit{number of adjectives}, \textit{number of adverbs}, \textit{number of adverbs that modify verbs}, \textit{number of cardinals}, and \textit{number of degree adjective/adverbs}. For each feature, we determine its rank based on the feature's probability of occurrence in the requirements of the data set. According 
%Corresponding
to Hussain et al.~\cite{hussain2008using}, we selected a cutoff threshold of $>0.8$. Therefore, we determined \textit{number of cardinals} and \textit{number of degree of adjectives/adverbs} as valid features among all five ones for the unprocessed data set. For the processed data set, we identified \textit{number of cardinals} and \textit{number of adverbs} as valid features.

Afterwards, we extract the required keyword features for the nine defined part-of-speech keyword groups \textit{adjective}, \textit{adverb}, \textit{modal}, \textit{determiner}, \textit{verb}, \textit{preposition}, \textit{singular noun}, and \textit{plural noun}. For each keyword group, we calculate the smoothed probability measure and selected the respective cutoff threshold manually to determine the most discriminating keywords for each data set, corresponding to Hussain et al. \cite{hussain2008using}.

Our final feature list for the unprocessed data set consisted of the ten features \textit{number of cardinals}, \textit{number of degree of adjectives/adverbs}, \textit{adjective}, \textit{adverb}, \textit{modal}, \textit{determiner}, \textit{verb}, \textit{preposition}, \textit{singular noun}, and \textit{plural noun}.

Our final feature list for the processed data set consisted of the ten features \textit{number of cardinals}, \textit{number of adverbs}, \textit{adjective}, \textit{adverb}, \textit{modal}, \textit{determiner}, \textit{verb}, \textit{preposition}, \textit{singular noun}, and \textit{plural noun}.

To classify each requirement of the respective data set, we implemented a Java-based feature extraction prototype that parses all requirements from the data set and extracts the values for all ten features mentioned above. Subsequently, we used Weka \cite{witten2016data} to train a C4.5 decision tree algorithm \cite{quinlan2014c4} which comes with Weka as J48 implementation. According to Hussain et al. \cite{hussain2008using}, we set the parameters for the minimum number of instances in a leaf to 6 to counter possible chances of over-fitting.

Since the data set was not very large with 625 requirements, we performed a 10-fold-cross validation. In the following, we report our classification results for each data set.\\
{\bf Results:} The classification of the \emph{unprocessed data set} results in $89.92\%$ correctly classified requirements with a weighted average precision and recall of $0.90$. The classification of the \emph{processed data set} results in $94.40\%$ correctly classified requirements with a weighted average precision of $0.95$ and recall of $0.94$. \tablename{ \ref{tb:classification_unprocessed}} and  \tablename{ \ref{tb:classification_processed}} show the details. By applying our approach, we could achieve an improvement of $4.48\%$ correctly classified requirements. In total, we could correctly classify $28$ additional requirements, which consist of $9$ functional and $19$ non-functional ones.
When classifying NFRs into sub-categories, the influence of our preprocessing is much stronger. The last two columns of \tablename{ \ref{tab:compare}} show the overall precision and recall of six different machine learning algorithms for sub-classifying NFRs into the categories listed in columns 1--10 of the table. For all algorithms, results are dramatically better when using the preprocessed data (column Total P) compared to using the raw data (column Total UP).

%\textit{(I) Unprocessed data set:} The classification of the unprocessed data set results in $89.92\%$ correctly classified requirements with an weighted average precision and recall of $0.9$. \tablename{ \ref{tb:classification_unprocessed}} shows the detailed classification results separated for functional and non-functional requirements.

\begin{table}[htbp]
	\centering
	\caption{Classification results of the unprocessed data set}

	\label{tb:classification_unprocessed}
	\resizebox{\linewidth}{!}{\begin{tabular}{c|c|c|c|c|c|c|}
		\cline{2-7}
		& \begin{tabular}[c]{@{}c@{}}Correctly \\ Classified\end{tabular} & \begin{tabular}[c]{@{}c@{}}Incorrectly \\ Classified\end{tabular} & Precision & Recall & F-Measure & Kappa \\ \hline
		\multicolumn{1}{|c|}{NFR} & 325 (87.84\%) & 45 (12.16\%) & 0.95 & 0.88 & 0.91 & \multirow{3}{*}{0.79} \\ \cline{1-6}
		\multicolumn{1}{|c|}{FR} & 237 (92.94\%) & 18 (7.06\%)& 0.84 & 0.93 & 0.88 &  \\ \cline{1-6}
		\multicolumn{1}{|c|}{Total} & 562 (89.92\%) & 63 (10.08\%) & 0.90 & 0.90 & 0.90 &  \\ \hline
	\end{tabular}}
\end{table}

%\textit{(II) Processed data set:}
%The classification of the processed data set results in $95.04\%$ correctly classified requirements with an weighted average precision and recall of $0.95$. \tablename{ \ref{tb:classification_processed}} shows the details. 

%detailed classification results separated for functional and non-functional requirements.

\begin{table}[htbp]
	\centering
	\caption{Classification results of the processed data set}

	\label{tb:classification_processed}
	\resizebox{\linewidth}{!}{\begin{tabular}{c|c|c|c|c|c|c|}
			\cline{2-7}
			& \begin{tabular}[c]{@{}c@{}}Correctly \\ Classified\end{tabular} & \begin{tabular}[c]{@{}c@{}}Incorrectly \\ Classified\end{tabular} & Precision & Recall & F-Measure & Kappa \\ \hline
			\multicolumn{1}{|c|}{NFR} & 344 (92.97\%) & 26 (7.03\%) & 0.98 & 0.93 & 0.95 & \multirow{3}{*}{0.89} \\ \cline{1-6}
			\multicolumn{1}{|c|}{FR} & 246 (96.47\%) & 9 (3.53\%)& 0.90 & 0.97 & 0.93 &  \\ \cline{1-6}
			\multicolumn{1}{|c|}{Total} & 590 (94.40\%) & 35 (5.60\%) & 0.95 & 0.94 & 0.94 &  \\ \hline
	\end{tabular}}

\end{table}

\vspace{-3mm}

%By applying our approach, we could achieve improvement of $5.12\%$ correctly classified requirements. In total, we could correctly classify additional $32$ requirements, which consist of $9$ functional and $23$ non-functional ones. 

\subsection{RQ2- Classifying Non-functional Requirements}
In this section, we describe the machine learning algorithms we used to classify NFRs. The performance of each method is assessed in terms of its recall and precision.
%under two main conditions: unprocessed and processed SRS  input documents.

%\begingroup
%\everymath{\scriptstyle}
%\scriptsize
%\[\small {\text{\it recall} = {\dfrac{TP}{TP+FN}}\quad \text{\it precision} = {\dfrac{TP}{TP+FP}}}\]\endgroup
%Where \(TP\), \(FN\), and \(FP\) represent True Positive, False Negative, and False Positive, respectively. 

\begin{figure*}[!ht]
\centering

\subfloat[{\scriptsize Hopkins statistic to assess the clusterability of the data set (hopkins-stat = 0.1)}]{\includegraphics[scale=0.26]{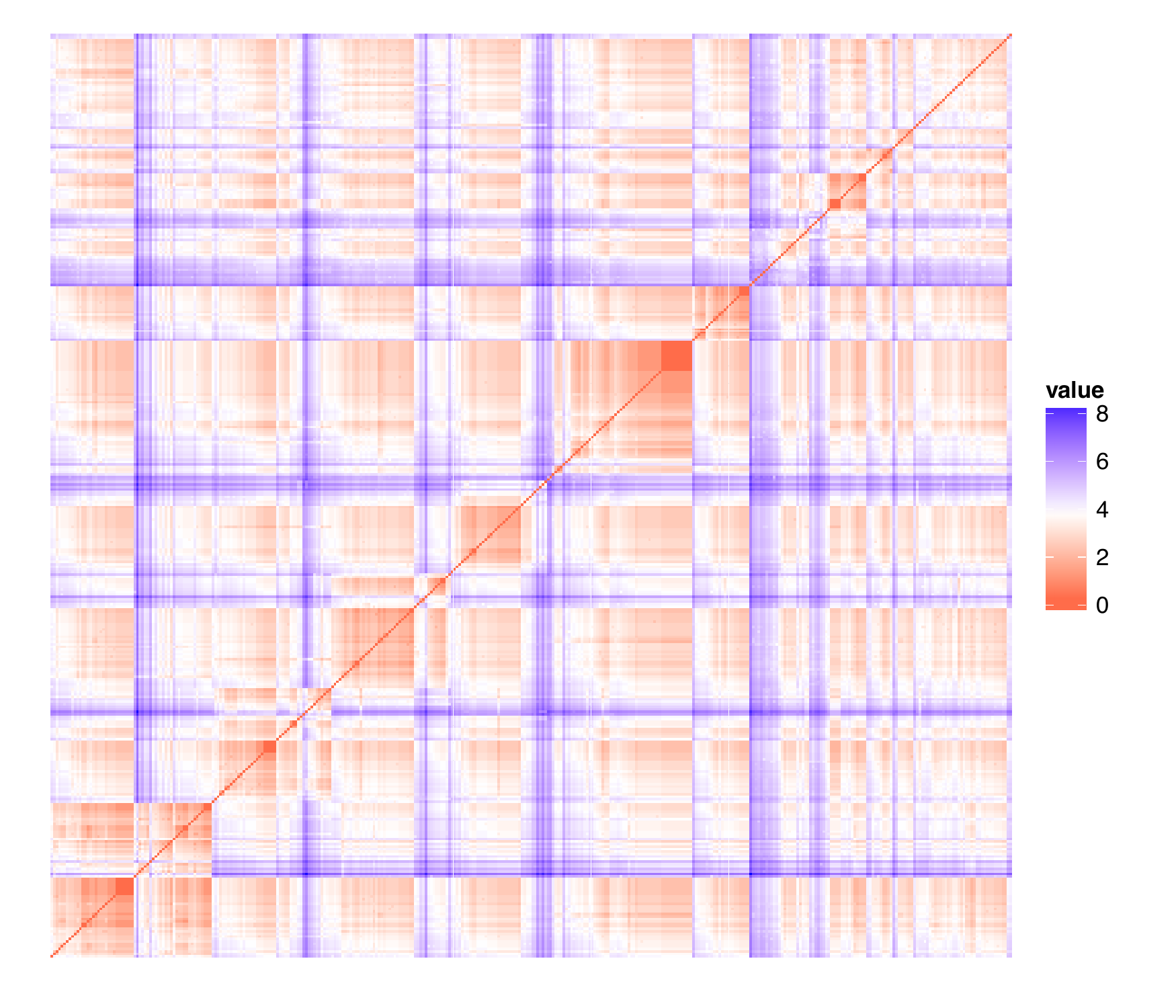}}
\subfloat[Hierarchical= 0.13]{\includegraphics[scale=0.28]{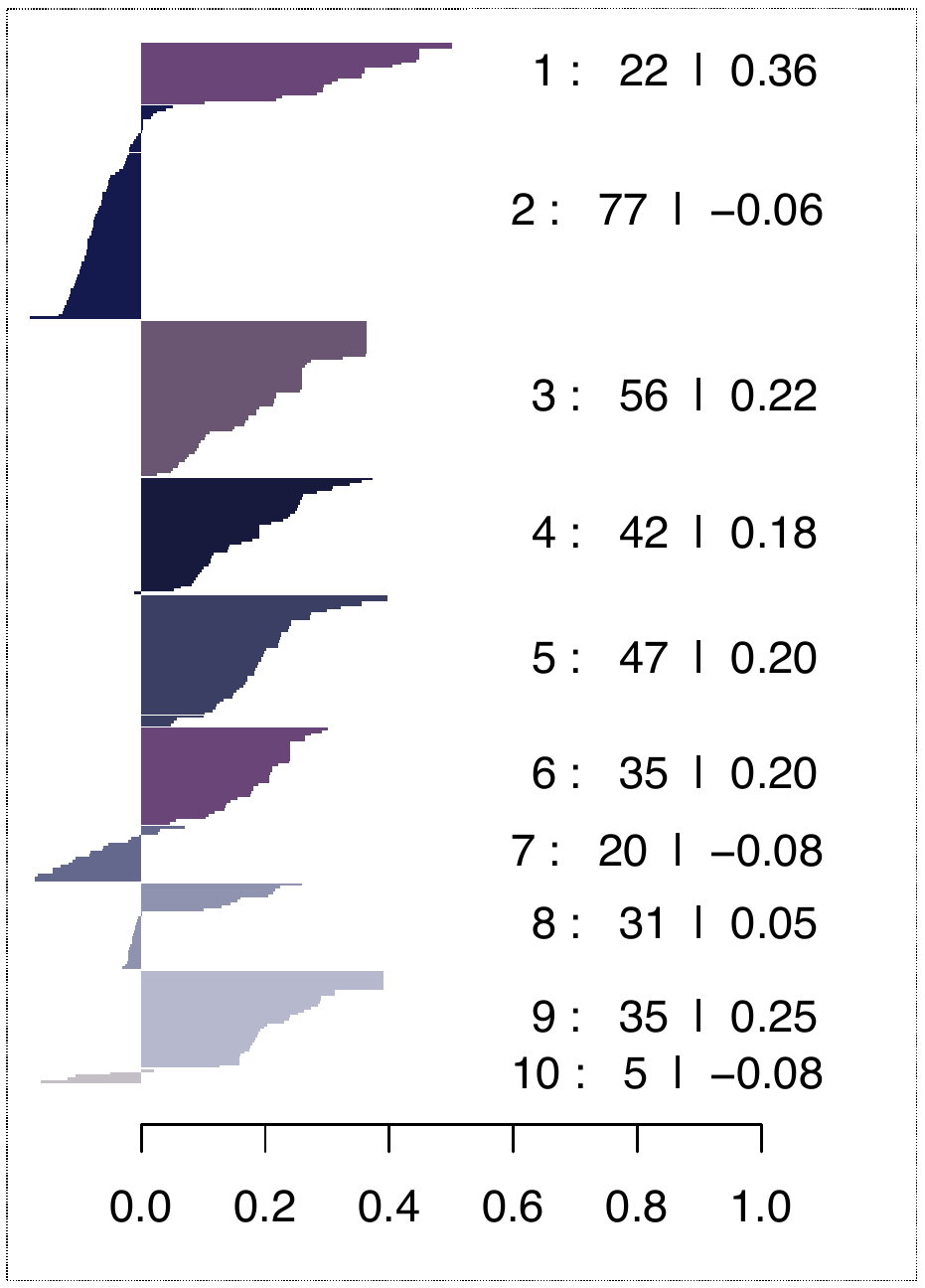}}
\subfloat[K-means= 0.1]{\includegraphics[scale=0.28]{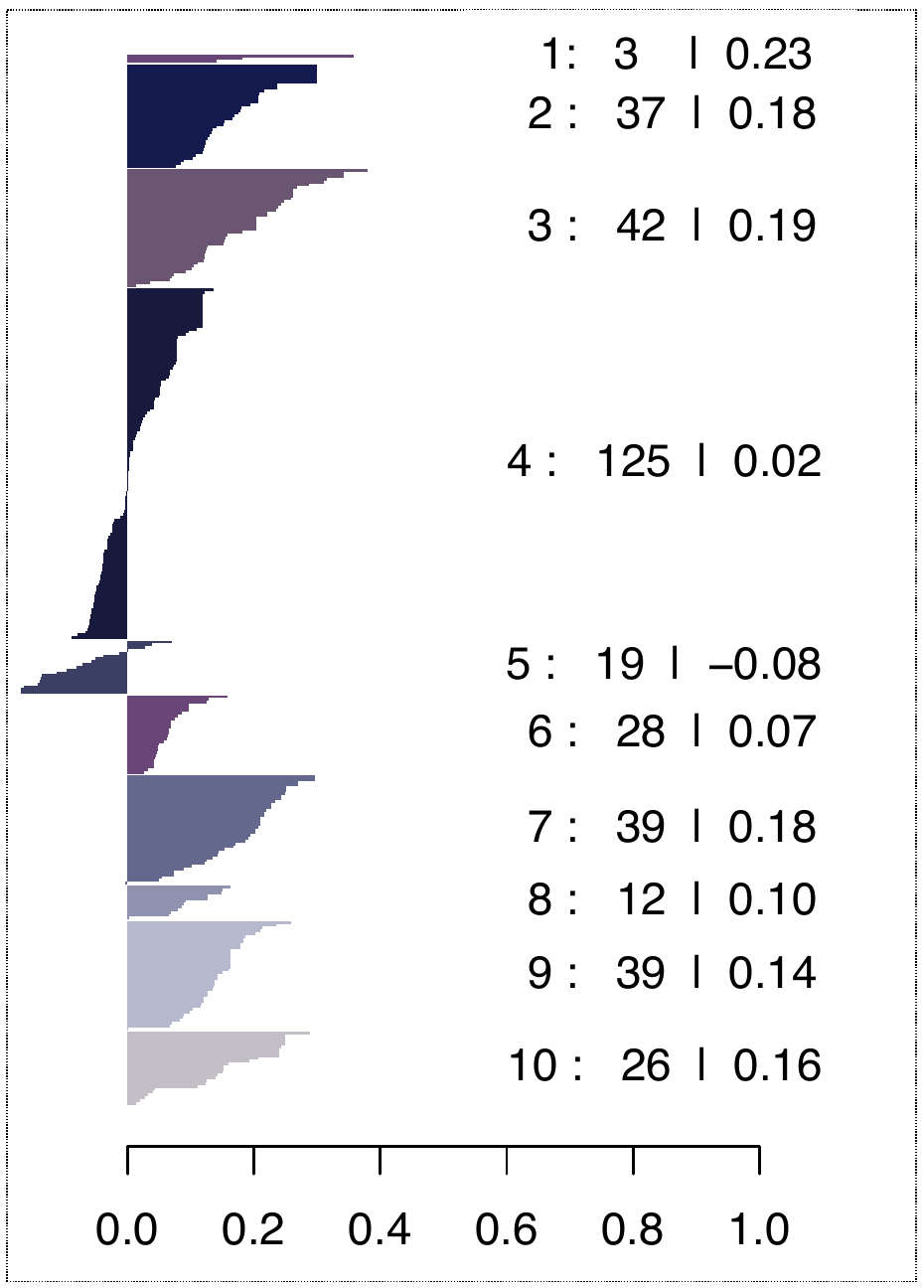}}
\subfloat[Hybrid= 0.13]{\includegraphics[scale=0.28]{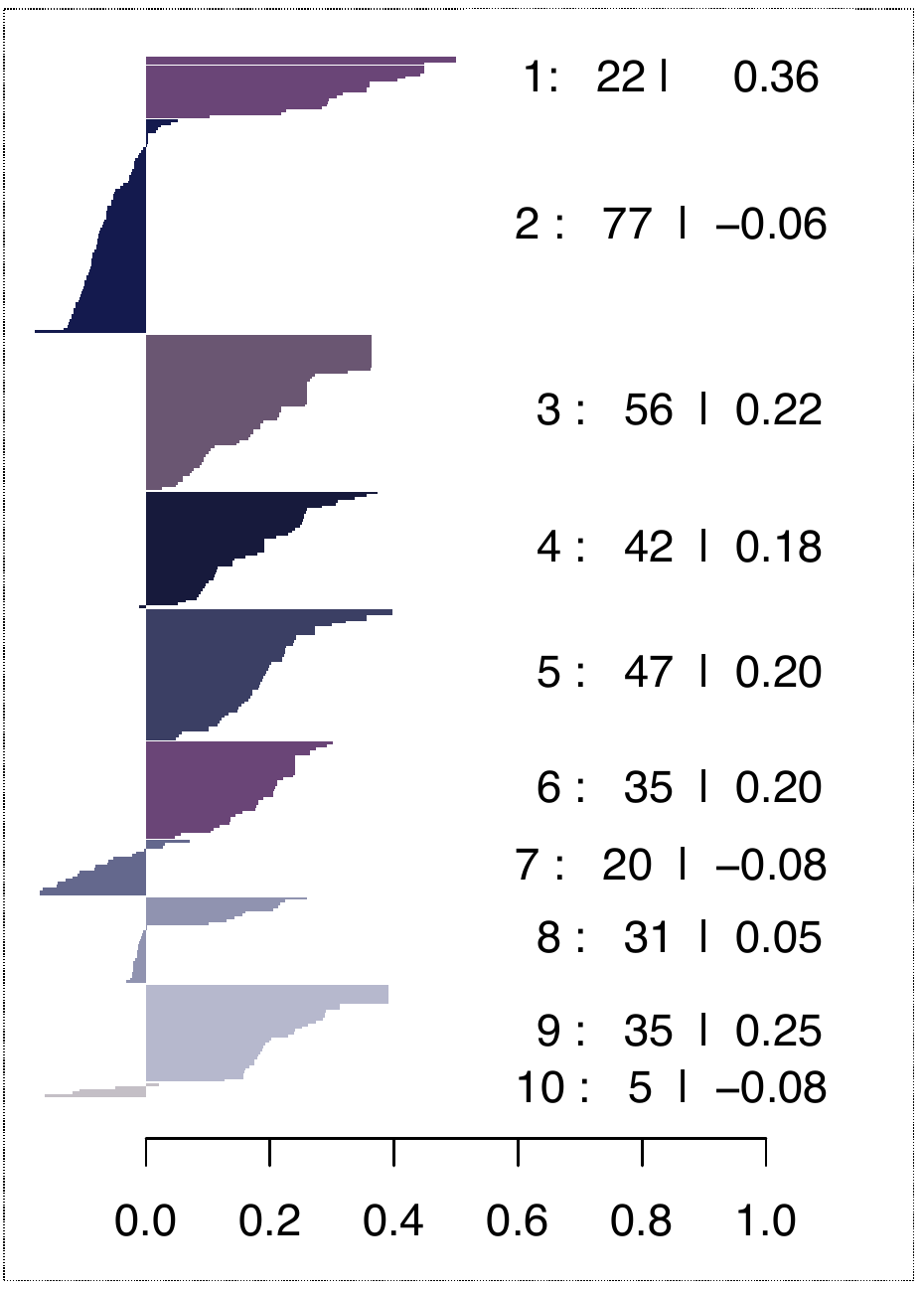}}
\subfloat[A visual representation of the confusion matrix (BNB algorithm)]{\includegraphics[scale=0.2]{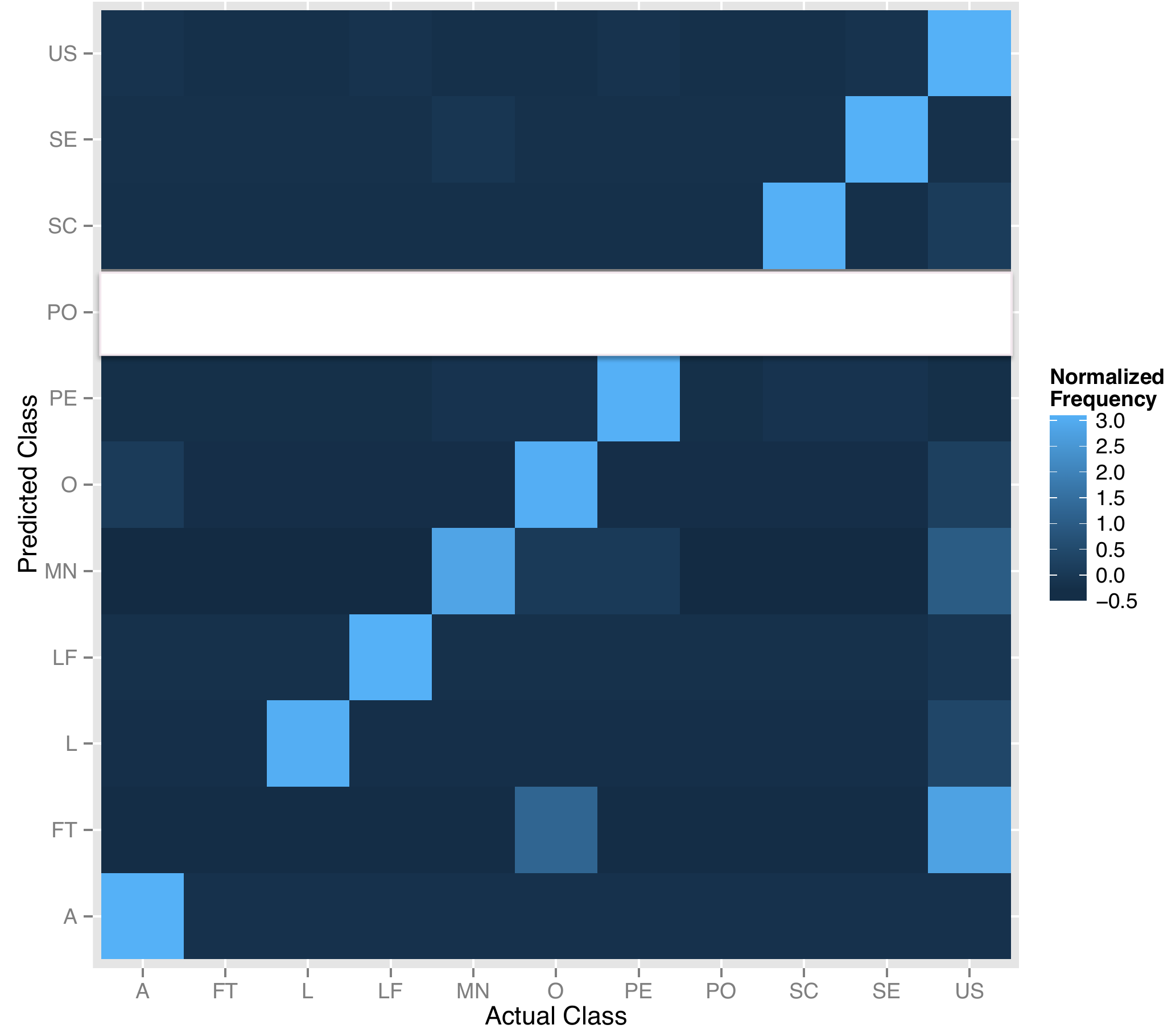}}\\
\caption{Detailed visual representation of classifying NFRs }
\label{fig:cluster}
\end{figure*}

\subsubsection{Topic Modeling}
Topic modeling is an unsupervised text analysis technique that groups a small number of highly correlated words, over a large volume of unlabelled text \cite{TM1}, into {\it topics}.

%\begin{figure}
%\centering
%\subfloat[LDA]{\includegraphics[scale=0.3]{Figures/LDA}}
%\subfloat[BTM]{\includegraphics[scale=0.3]{Figures/BTM}}
%\caption{The graphical model of LDA and BTM approaches in plate notation}
%\end{figure}
{\bf Algorithms:} The {\it Latent Dirichlet Allocation (LDA)} algorithm classify documents based on the frequency of word co-occurrences. Unlike the LDA approach, the {\it Biterm Topic Model} (BTM) method models topics based on the word co-occurrence patterns and learns topics by exploring word-word (i.e., biterm) patterns. Some recent studies on the application of topic modeling in classifying short text documents stated that the BTM approach has a better ability in modeling short and sparse text, as the ones typical for requirements specifications.

\begin{table*}
\centering
\footnotesize
\caption{Comparison between classification algorithms for classifying non-functional requirements [(U)P= (Un)Processed]}
\label{tab:compare}

\begin{tabular}{ |c|cc|cc|cc|cc|cc|cc|cc|cc|cc|cc|cc|cc| cc| }
  \hline
{\bf Algorithm}& \multicolumn{2}{c|}{\bf A}& \multicolumn{2}{c|}{\bf US}& \multicolumn{2}{c|}{\bf SE}& \multicolumn{2}{c|}{\bf SC}& \multicolumn{2}{c|}{\bf LF}& \multicolumn{2}{c|}{\bf L}& \multicolumn{2}{c|}{\bf MN}& \multicolumn{2}{c|}{\bf FT}& \multicolumn{2}{c|}{\bf O}& \multicolumn{2}{c|}{\bf PE} &  \multicolumn{2}{c|}{\bf PO}& \multicolumn{2}{c|}{\bf Total [P]}& \multicolumn{2}{c|}{\bf Total [UP]}\\ \cline{2-27}
&{\bf R}&{\bf P}   &{\bf R}&{\bf P}   & {\bf R}&{\bf P}&   {\bf R}&{\bf P}   &{\bf R}&{\bf P}   &{\bf R}&{\bf P}   &{\bf R}&{\bf P}   &{\bf R}&{\bf P}   &{\bf R}&{\bf P}   &{\bf R}&{\bf P} & {\bf R}&{\bf P} & {\bf R}&{\bf P}&{\bf R}&{\bf P}\\\hline
{\bf LDA}&95&60   &61&76   & 87&87&   81&57   &60&85   &47&20   &70&52   &10&2   &35&70   &70&95 & -&-   &\cellcolor{Gray}62&\cellcolor{Gray}62&31&31\\\cline{2-27}
{\bf BTM}&0&0   &6&12   & 13&18&   9&8   &5&7   &0&0   &0&0   &40&17   &0&0   &18&43 & -&-   &\cellcolor{Gray}8&\cellcolor{Gray}8&3&3\\\cline{2-27}
%{\bf BTM} &&&&&&&&&&&&&&&&&&&&&&&&\\ \cline{2-25}
{\bf Hierarchical }&13&14   &25&20   &24&17   &16&29   &5&3   &6&15   &19&35   &18&40   &32&29   &26&22   &-&-  & \cellcolor{Gray}21&\cellcolor{Gray}21&16&16\\ \cline{2-27}
{\bf K-means}&10&23   &19&14   &29&18   &14&14   &21&21   &8&15   &22&47   &18&40   &26&30   &31&11&   -&-   &\cellcolor{Gray}20&\cellcolor{Gray}20&15&15\\ \cline{2-27}
{\bf Hybrid}&15&14   &27&22&   29&18&   20&4&   26&24   &6&15   &17&35   &18&40   &22&27   &26&22&   -&-   &\cellcolor{Gray}22&\cellcolor{Gray}22&19&17\\\cline{2-27}
{\bf Na\"{i}ve Bayes} &\cellcolor{Gray}90&\cellcolor{Gray}90&\cellcolor{Gray}   97&\cellcolor{Gray}77   &\cellcolor{Gray}97&\cellcolor{Gray}100   &\cellcolor{Gray}83&\cellcolor{Gray}83   &\cellcolor{Gray}94&\cellcolor{Gray}94   &\cellcolor{Gray}75&\cellcolor{Gray}100   &\cellcolor{Gray}90&\cellcolor{Gray}82   &\cellcolor{Gray}97&\cellcolor{Gray}90   &\cellcolor{Gray}78&\cellcolor{Gray}91   &\cellcolor{Gray}90&\cellcolor{Gray}100   &-&-   &\cellcolor{Gray}91&\cellcolor{Gray}90&45&45\\\hline
\end{tabular}

\end{table*}
{\bf Results and Evaluation:} The modeled topics for both LDA and BTM, including the top frequent words and the NFR assigned to each topic are provided in our source code package\footnote{http://wcm.ucalgary.ca/zshakeri/projects}. We determined each topic by the most probable words that are assigned to it. For instance, LDA yields the word set \{user, access, allow, prior, and detail\} for the topic describing the Fault Tolerance sub-category, while BTM yields the set \{failure, tolerance, case, use and data\}.

Generally, the word lists generated by BTM for each topic are more intuitive than those produced by LDA. This confirms previous research that BTM performs better than LDA in terms of modeling and generating the topics/themes of a corpus consisting of short texts. However, surprisingly, BTM performed much worse than LDA for sub-classifying NFRs as shown in Table~\ref{tab:compare}. This might be because BTM performs its modeling directly at the corpus level and biterms are generated independently from topics. 
%However, the word lists generate by BTM for each topic were more intuitive than LDA (e.g. generated words for the fault tolerance NFR in LDA are not directly this topic). These implies that, the BTM approach generally performs better than LAD in terms of modelling and generating the topics/themes of a short texts corpus. However it performed very poor in assigning the documents (e.g., NFRs) to topics.

%--------------------------------------------------------------------------
 \PRLsep 
  \vspace{-3mm}
\subsubsection{Clustering}
Clustering is an unsupervised classification technique which categorizes documents into groups based on likeness \cite{Cluster}. This likeness can be defined as the numerical distance between two documents \(D_i\) and \(D_j\) which is measured as:
\vspace{-3mm}

\begingroup
\everymath{\scriptstyle}
\scriptsize
\[
d(D_i, D_j)= \sqrt{({d_i}_1-{d_j}_1)^2+ ({d_i}_2-{d_j}_2)^2 + ...+ ({d_i}_n-{d_j}_n)^2}
\] \endgroup

Where \(({d_i}_1, {d_i}_2, ..., {d_i}_n)\) and \(({d_j}_1, {d_j}_2, ..., {d_j}_n)\) represent the coordinates (i.e., word frequencies) of the two documents.

{\bf Algorithms: }The {\it Hierarchical} (i.e., Agglomerative) algorithm, first, assigns each document to its own cluster and iteratively merges clusters that are closest to each other until the entire corpus forms a single cluster. Despite the hierarchical approach in which we do not need to specify the number of clusters upfront, the {\it K-means} algorithm assigns documents randomly to {\it k} bins. This approach computes the location of the centroid of each bin and computes the distance between each document and each centroid. We defined the {\it k=10} to run this algorithm. However, the k-means approach is highly sensitive to the initial random selection of cluster centroid (i.e., mean), which might lead to different results each time we run this algorithm. Thus, we used a {\it Hybrid} algorithm, which combined hierarchical and K-means algorithms. This algorithm, first, computes the center (i.e., mean) of each cluster by applying the hierarchical approach. Then computes the K-means approach by using the set of defined clusters' centers.

{\bf Results and Evaluation: }Before applying clustering algorithms we used Hopkins (H) statistic to test the spatial randomness and assess the {\it clustering tendency} (i.e., clusterability) of our data set. To this end, we raised the following null hypothesis: \(\big(\){\it \(H_0\): the NFR data set is uniformly distributed and has no meaningful clusters}\(\big)\). As presented in Figure \ref{fig:cluster} (a), the {\it H-value} of this test is 0.1 (close to zero), which rejects this hypothesis and concludes that our data set is significantly clusterable. However, as presented in Table \ref{tab:compare}, the clustering algorithms had poor performance at classifying NFRs. This may imply that the data set under study is quite unstructured and sub-categories of NFRs are not well separated. Thus, an unsupervised algorithm (e.g. Hierarchical or K-means) cannot accurately achieve segmentation.

Moreover, we used Silhouette (s) analysis to assess the cohesion of resulted clusters. We used the function {\it silhouette()} of {\it cluster package} to compute the silhouette coefficient. Small {\it s-value} (i.e., around 0) means that the observation lies between two clusters and has a low cohesion. The results of this test and the details of each cluster, including a number of requirements assigned to it, and its {\it s-value} are illustrated in Figure \ref{fig:cluster}(b-d). 

 \PRLsep 
  \vspace{-3mm}
  
\subsubsection{Na\"{i}ve Bayes Classification}

This approach is a supervised learning method which predicts unseen data based on the {\it bayes' theorem} \cite{Naive} used to calculate conditional probability:

\begingroup
\everymath{\scriptstyle}
\scriptsize
\[
P(C=c_k\mid F=f)= \dfrac{P(F=f\mid C=c_k) P(C=c_k)}{P(f)}
\] \endgroup

Where \(C=(c_1, c_2, ..., c_k)\) represents classes and \(F= (f_1, f_2, ..., f_d)\) is a vector random variable, which includes one vector for each document.

{\bf Algorithm:} We use a variation of the multinomial Na\"{i}ve Bayes (BNB) algorithm known as {\it Binarized Na\"{i}ve Bayes}. In this method, the term frequencies are replaced by Boolean presence/absence features. The logic behind this is the higher importance of word occurrence than word frequency to sentiment classification.
%The logic behind this is that for sentiment classification, word occurrence matters more than word frequency.

{\bf Results and Evaluation:} To apply this algorithm we employed a 5-fold-cross validation. To reduce the data splitting bias, we run five runs of the 5-fold-cross validation. Overall accuracy is calculated at just over 90\% with a {\it p-value} of 2.2e-16. As illustrated in Table \ref{tab:compare}, results obtained using the BNB algorithm were generally more accurate. All of the NFRs (except for PO) were recalled at relatively high values ranging from 75 (i.e., Legal requirements) to 97\% (i.e., security and performance requirements). To represent more details about the performance of our classifier for each NFR, we visualized the confusion matrix resulted from applying the BNB algorithm (Figure \ref{fig:cluster} (e)). Each column and row of this matrix represent the actual (i.e., reference) and the prediction data, respectively. The blocks are colored based on the frequency of the intersection between actual and predicted classes (e.g., the diagonal represents the correct predictions for the actual class). Since some of the NFRs in our data set occur more frequently, we normalized our data set before visualizing the confusion matrix. As illustrated in Figure \ref{fig:cluster} (e), requirements in classes FT, L, MN, O, and SC were often assigned to class US. We can imply that {\it the terminology we use for representing usability requirements is very general, which covers other NFRs that are indirectly related to usability. } This shows a clear need for additional (or better) sentimental patterns, which differentiate this category of NFR from other similar categories. 
to differentiate usability requirements from other types of NFRs.

\begin{tcolorbox}[colback=white, title= Findings]
\footnotesize
  \textcolor{white}{..................  }  \\
\vspace{-5mm}
    
{\bf Finding 1:} Our preprocessing approach positively impacted the performance of the applied classification of functional and non-functional requirements. We could improve the accuracy from 89.92\% to 95.04\%.\\
\vspace{-2mm}

{\bf Finding 2:} Our preprocessing approach strongly impacted the performance of all applied sub-classification methods. For LDA and BNB, both precision and recall doubled. \\
\vspace{-2mm}

{\bf Finding 3:} Among the  machine learning algorithms LDA, BTM, Hierarchical, K-means, Hybrid and Binarized Na\"{i}ve Bayes (BNB), {\it BNB} had the highest performance for sub-classifying NFRs.\\
\vspace{-2mm}

{\bf Finding 4:} While BTM generally works better than LDA for exploring the general themes and topics of a short-texts corpus, it did not perform well for sub-classifying NFRs. \\
\vspace{-2mm}

{\bf Finding 5:} There is a clear need for additional sentimental patterns/sentence structures to differentiate usability requirements from other types of NFRs.
\end{tcolorbox}
\vspace{-2mm}

%\begin{figure}
%\centering
%{\includegraphics[scale=0.35]{Figures/confusion}}\\
%\caption{A visual representation of the confusion matrix of the binarized Na\"{i}ve Bayes algorithm}
%\label{fig:naive}
%\end{figure}

\vspace{-1mm}
\section{Limitations and Threats to Validity}
\label{sec:limits}
In this section, we discuss the potential threats to the validity
of our findings in two main threads: 

{\bf (1) Data Analysis Limitations:} The biggest threat to the validity of this work is the fact that our preprocessing model was developed on the basis of the data set given for the RE Data Challenge and that we had to use the same data set for evaluating our approach.
%One of the important limitations of this work is that the performance of the applied techniques is biased by our pre-processing model and the data set we used for proposing this approach. 
We mitigate this threat by using sentence structure features such as temporal, entity, and functional features which are applicable to sentences with different structures from different contexts. We also created a set of regular expressions which are less context-dependent and have been formed mainly based on the semantics of NFRs.

Another limiting factor is that our work depends on
%The presented work is also  limited by 
the number and choice of the NFR sub-categories used by the creators of our data set.
%which corresponds to the number of frequent NFRs we used for our data analysis and evaluation. 
However, our preprocessing can be adapted to a different set of NFR sub-categories. In terms of the co-occurrence rules and regular expressions presented in Table \ref{tab:reg}, we aim to
expand these rules by adding more NFR sub-categories in future work, as we gain additional insights from processing real world requirements specifications.

{\bf (2) Dataset Limitations:} Due to the nature of the RE'17 Data Challenge, we used the data set as is, although it has major data quality issues: (1)~Some requirements are incorrectly labeled. For example, R2.18 ``The product shall allow the user to view previously downloaded search results, CMA reports and appointments'' is labeled as NFR. Obviously, however, this is a functional requirement.
%Requirement 2.26 ``The product is expected to run on Windows CE and Palm operating systems'', labeled as PO, is not a portability requirement, but a technical constraint.
(2) The important distinction between quality requirements and constraints is not properly reflected in the labeling. (3)~The selection of the requirements has some bias. For example, the data set does not contain any compatibility, compliance or safety requirements. Neither does it contain any cultural, environmental or physical constraints. (4)~Only one single requirement is classified as PO which makes this sub-category useless for our study. The repetition of our study on a data set of higher data quality is subject to future work.

Furthermore, the {\it unbalanced data set} we used for classifying the NFRs may affect the findings of this study. However, a study by Xue and Titterington \cite{unbalanced} revealed that there is no reliable empirical
evidence to support the claim that an unbalanced data
set negatively impacts the performance of the LDA/BTM approaches. Further, a recent study by L\'{o}pez et al. \cite{unbalanced2} shows that the unbalanced ratio by itself does not have the most significant effect on the classifiers’
performance, but there are other issues such as (a) the presence of small disjuncts, (b) the lack of density, (c) the class overlapping,
(d) the noisy data, (e) the management of borderline examples, and (f) the dataset shift that must be taken into account. The pre-processing step we conducted before applying the classification algorithms helps discriminate the NFR sub-classes more precisely and mitigates the negative impact of the noisy data and borderline problems. Moreover, we employed the n-fold cross validation technique which helps generate enough positive class instances in different folds and reduces additional problems in the data distribution especially for highly unbalanced datasets. This technique, to a great extent, mitigates the negative impact of class overlapping, the dataset shift, and the presence of small disjuncts issues on the performance of the classification algorithms we applied in this study.

\section{Conclusion and Implications}
\label{sec:agenda}

Our findings are summarized in the box at the end of Section~\ref{sec:results}. 
In particular, we conclude that using our preprocessing approach improves the performance of both classifying FR/NFR and sub-classifying NFR into sub-categories. Further, we found that, among popular machine learning algorithms, Binarized Na\"{i}ve Bayes (BNB) performed best for the task of classifying NFR into sub-categories. Our results further show that, although BTM generally works better than LDA for extracting the topics of short-texts, BTM does not perform well for classifying NFRs into sub-categories. Finally, additional (or better) sentimental patterns and sentence structures are needed for differentiating usability requirements from other types of NFRs.
%MG: from where do you conclude this result?? This research showed that sentimental patterns and sentence structures have the potential to be used for differentiating usability requirements from other types of NFRs.
\vspace{-2mm}

%
%\begin{algorithm}
% \KwIn{words \(w \in\) documents \(d\), \(nstart\), \(burn-in\), \(thin\), \(n_{d,k}\)}
% \KwOut{topic assignments \(z\)}
% \Begin{
% randomly initialize \(z\) and increment counters 
% {
%
%  \ForEach{\(iteration\)}{
%  \ForEach {\(word\) \(w\)} 
%  {
%  \ForEach {\(topic\) \(k\)}{
%  \(\theta_{d_{w},k}\) = {\it calculating the document/topic distribution for topic \(k\), word \(w\) in document \(d\)}
%  
%  }
%  \(topic\gets\) sample from \(multnomial (\theta_{d_{w}})\)\\
%  \(z[w]\gets topic\)\\
%  {\it update counts according to new assignments}
%  }
%  }
%   \Return{\(z\)}
% }
% }
% \caption{LDA Gibbs Algorithm \cite{algorithm} for exploring the most popular on Q\&A sites}
% \label{alg:alg}
% 
%\end{algorithm}

% \fi

%\footnotesize
\bibliographystyle{IEEEtran}
\bibliography{IEEEabrv,refs}

\end{document}